\documentclass[prd,onecolumn,superscriptaddress]{revtex4}
\usepackage{graphicx}
\usepackage{amsmath}
\usepackage{amsfonts}
\usepackage{amssymb}
\usepackage{color}
\usepackage{amscd}
\usepackage{bm}
\makeatletter
\renewcommand{\maketag@@@}[1]{\hbox{\m@th\normalsize\normalfont#1}}%
\makeatother
\begin{document}
\title{Logarithmic correction to black hole entropy from the nonlocality of quantum gravity}
\author{Yong Xiao}
\email{xiaoyong@hbu.edu.cn}
\affiliation{Key Laboratory of High-precision Computation and Application of Quantum Field Theory of Hebei Province,
College of Physical Science and Technology, Hebei University, Baoding 071002, China}
\author{Yu Tian}
\email{ytian@ucas.ac.cn}
\affiliation{School of Physics, University of Chinese Academy of Sciences, Beijing 100049, China}
\affiliation{Institute of Theoretical Physics, Chinese Academy of Sciences, Beijing 100190, China}

\begin{abstract}
It has been known for many years that the leading correction to the black hole entropy is a logarithmic term, which is universal and closely related to conformal anomaly. A fully consistent analysis of this issue has to take quantum backreactions to the black hole geometry into account. However, it was always unclear how to naturally derive the modified black hole metric especially from an effective action, because the problem refers to the elusive non-locality of quantum gravity. In this paper, we show that this problem can be resolved within an effective field theory (EFT) framework of quantum gravity. Our work suggests that the EFT approach provides a powerful and self-consistent tool for studying the quantum gravitational corrections to black hole geometries and thermodynamics.
\end{abstract}
 \maketitle

\section{Introduction}

The study of black hole entropy plays a crucial role in the understanding of the microscopic degrees of freedom of quantum gravity. The famous Bekenstein--Hawking entropy takes the form $S_{BH}= \frac{\mathcal{A}}{4}$. From a more general viewpoint of quantum gravity, this entropy is only considered as the tree level result of the black hole entropy \cite{Maldacena:1997de,Solodukhin:2011gn}, and the leading order quantum gravitational correction to the  entropy has been shown to be a logarithmic term. The logarithmic correction is fundamental and universal, since it can be derived from many different approaches such as conical singularity and entanglement entropy \cite{Solodukhin:2011gn,Solodukhin:1994yz}, Euclidean action method \cite{Fursaev:1994te,Sen:2012dw,El-Menoufi:2015cqw,El-Menoufi:2017kew}, conformal anomaly \cite{Cai:2009ua}, Cardy formula \cite{Carlip:2000nv}, quantum tunneling \cite{Banerjee:2008cf,Banerjee:2008fz}, and quantum geometry \cite{kaul}. See also the references therein.

A rapid dimensional analysis is sufficient to uncover this logarithmic term of the entropy. Consider the leading order quantum gravitational corrections to the thermodynamic quantities of a Schwarzschild black hole. Since the first order corrections should have an extra factor $\hbar G$, we expect the modified black hole temperature be written as
$T=\frac{1}{8\pi M} (1 -   \hbar G \frac{ a_1 }{ M^2})$, where the $\frac{1}{M^2}$ scaling behavior is to balance the dimension of  $\hbar G$, and the coefficient $a_1$ is to be fixed by a more fundamental analysis. Due to the thermodynamic law $Td S=d M$, it immediately leads to the entropy $S=4\pi M^2  +4\pi a_1 \hbar G \ln (M^2)$ up to an integration constant. In fact, in the seminal paper by Fursaev \cite{Fursaev:1994te}, the one-loop corrections to the thermodynamics of a Schwarzschild black hole have been analyzed, and the results are
\begin{align}
T=\frac{1}{8\pi M}-\gamma\frac{2}{M^3},\ \ \ S=4\pi M^2 +64 \pi^2 \gamma \ln(M^2),\label{cther}
\end{align}
where we set $G=\hbar=c=1$ for simplicity, and the coefficient $\gamma$ is calculable and defined as eq.\eqref{gamma}.

However, the backreaction of the quantum effects to the back hole geometry had not been taken into account. In fact, the corrections to the thermodynamics must imply the original Schwarzschild geometry also receives quantum corrections. For example, the black hole temperature should be related to the conical singularity of the metric; the temperature of the form \eqref{cther} can not be consistent with the original Schwarzschild metric. Fursaev stressed that it would be incomplete without obtaining these thermodynamic behaviors directly from the modified geometry \cite{Fursaev:1994te}.

Despite its significance, this question remained unanswered over decades especially from the level of effective action, and it appeared to be quite difficult and unlikely to find any analytic solution \cite{El-Menoufi:2015cqw,Solodukhin:2019xwx}. Obviously, from the Wald entropy formula, any possible higher order curvature terms of a \emph{local} form that can be added to the Einstein-Hilbert action are unable to produce such a logarithmic term. So the logarithmic correction has to be related to some kinds of \emph{non-local} effects of quantum gravity, as was alluded to in the literature \cite{Solodukhin:1997yy,Cai:2009ua}.

 In this paper, we will show the problem can be resolved in the framework of EFT of gravity. The paper is organized as follows. First, we display the one-loop effective action of quantum gravity and show how the trace anomaly is illustrated in this formalism. Then, we show that the non-local terms in the effective action produce both the quantum corrections to the geometry and the thermodynamic quantities such as the temperature and entropy of the Schwarzschild black hole, which resolves the above mentioned problem. Our approach can be easily generalized to other types of black hole. We will provide the results for the Reissner-Nordstr{\"o}m (R-N) black hole and AdS-Schwarzschild black hole. Everything fits together nicely even in the complicated cases, manifesting the self-consistencies of our analysis.

\section{The one-loop effective action of quantum gravity and trace anomaly} \label{sec2}

In the EFT of gravity, starting from the action $I= \int_{\mathcal{M}} \frac{1}{16 \pi} R + I_{m}$ and integrating out the quantum fluctuations of matter and graviton at one-loop level, one obtains the effective action of quantum gravity. At second order in curvature, there is \cite{Donoghue:1994dn,Donoghue:2014yha,Barvinsky:1985an,Calmet:2019eof,Teixeira:2020kew}
\begin{align}
\begin{split}
I_{\text{EFT}} &= \int_{\mathcal{M}} \frac{R}{16\pi} +\left[ c_1(\mu) R^2 + c_2(\mu) R_{\mu\nu} R^{\mu\nu} + c_3(\mu) R_{\mu\nu\rho\sigma} R^{\mu\nu\rho\sigma} \right] \\
&-\left[\alpha R \ln(\frac{\Box}{\mu^2} )R + \beta R_{\mu\nu} \ln(\frac{\Box}{\mu^2}) R^{\mu\nu} + \gamma R_{\mu\nu\alpha\beta} \ln( \frac{\Box}{\mu^2} ) R^{\mu\nu\alpha\beta} \right],
\end{split}\label{EFTaction}
\end{align}
where $\Box \equiv -g^{\mu\nu} \nabla_\mu \nabla_\nu$ with the metric signature $(-1,1,1,1)$. The first part is the classical Einstein-Hilbert action, while the other part represents quantum effects. The non-local operator $\ln \Box$ is generated from the loop fluctuations of the massless particles, and the coefficients $\alpha$, $\beta$ and $\gamma$ are calculable and given as \cite{qfcs,Donoghue:2014yha}
\begin{align}
    \alpha &= \frac{1}{11520\pi^2}(5 (6\xi-1)^2 N_s-5 N_f -50 N_v+430 N_g),\label{a1} \\
    \beta &=\frac{1}{11520\pi^2}(-2 Ns + 8 N_f+176 N_v -1444 N_g), \label{a2}\\
    \gamma &=\frac{1}{11520\pi^2} (2N_s + 7 N_f -26 N_v +424 N_g ), \label{gamma}
\end{align}
where $N_s$, $N_f$, $N_v$ and $N_g$ are the number of scalars, four-component fermions, vectors and gravitons in the low energy particle spectrum in nature. These coefficients in eq.\eqref{EFTaction} represent the model-independent prediction of EFT, and should be obeyed by any candidate of a complete quantum gravitational theory if only it admits the same particle spectrum at low energy limit. Notice the formalism of EFT has already been renormalization group invariant, due to the property of the beta function of the Wilson coefficients such as $c_3(\mu)=c_3(u_*)-\gamma \ln(\frac{\mu^2}{\mu_*^2})$ etc \cite{El-Menoufi:2015cqw}.

The quantum vacuum is expected to be sightly shifted away from the classical vacuum. Especially, after considering the quantum fluctuations of the vacuum, the Einstein tensor is not necessarily traceless. This phenomenon is called trace anomaly and is transparent from the perspective of effective action \eqref{EFTaction}. We abbreviate the effective action \eqref{EFTaction} as the form $I_{\text{EFT} }=I_{\text{EH} } +  I_q$. The Einstein tensor comes from the variation of the classical Einstein-Hilbert part $\delta I_{\text{EH} }$, which vanishes at the classical vacuum. However, when the action is modified by quantum effects, we should have $\delta I_{\text{EFT} }= \delta I_{\text{EH} } + \delta  I_q=0$, and generally the two parts do not vanishes independently but cancel each other. Thus, to extract the value of the trace of the Einstein tensor, we only need to study $\delta I_q= \frac{1}{2} \int d^4 x \sqrt{-g}T^{\mu\nu}\delta g_{\mu\nu}$ in detail, where $T^{\mu\nu}$ denotes the effective energy-momentum tensor caused by quantum fluctuations. Considering a special variation of $I_q$ with respect to $g_{\mu\nu} \rightarrow e^{\epsilon} g_{\mu\nu}$, it becomes
\begin{align}
\delta I_q= \frac{1}{2} \int \sqrt{g}T^{\mu}_{\ \mu}\epsilon. \label{vari}
\end{align}
 Concretely, due to $ \delta \ln \Box= \ln  \Box'-\ln  \Box=- \epsilon$, the variation $\delta I_q$ becomes $\alpha R (\delta \ln\Box) R+ \beta R_{\mu\nu} (\delta \ln\Box) R^{\mu\nu} + \gamma R_{\mu\nu\alpha\beta}  (\delta \ln\Box) R^{\mu\nu\alpha\beta}= -(\alpha R^2 + \beta R_{\mu\nu} R^{\mu\nu} + \gamma R_{\mu\nu\alpha\beta}  R^{\mu\nu\alpha\beta})\epsilon$. It gives the trace anomaly
\begin{align}
  T^\mu_{\ \mu}= 2 (\alpha R^2 + \beta R_{\mu\nu} R^{\mu\nu} + \gamma R_{\mu\nu\alpha\beta}  R^{\mu\nu\alpha\beta}).\label{anom}
 \end{align}
 In the above analysis, we temporarily omit the terms like $\gamma (\delta R_{\mu\nu\alpha\beta}) \ln \Box R^{\mu\nu\alpha\beta}$ in $\delta I_q$, which may also contribute to $T^\mu_{\ \mu}$. However, if variating the action around Ricci-flat geometries, such as the original Schwarzschild metric, these contributions vanish, as will see shortly.

We emphasize that, for the conformal fields, the well-known effect of conformal anomaly has been contained above. Taking $\xi=\frac{1}{6}$ for the conformally coupled scalars and ignoring the gravitons in eqs.\eqref{a1}-\eqref{gamma}, the coefficients $\alpha$, $\beta$ and $\gamma$ are not independent with each other, then eq.\eqref{anom} can be rewritten as the standard form of the conformal anomaly \cite{Duff:2020dqb}
\begin{align}
\begin{split}
 T^\mu_{\ \mu}= \lambda_1 F  - \lambda_2 G,\label{cfano}
 \end{split}
\end{align}
where the two bases are the square of Weyl curvature $F=C_{\mu\nu\rho\sigma}C^{\mu\nu\rho\sigma}=\frac{1}{3} R^2 -2 R_{\mu\nu} R^{\mu\nu} + R_{\mu\nu\rho\sigma} R^{\mu\nu\rho\sigma}$, and the Euler density $G=R^2 -4 R_{\mu\nu} R^{\mu\nu} +  R_{\mu\nu\rho\sigma} R^{\mu\nu\rho\sigma}$, with the combination coefficients
\begin{align}
  \lambda_1=\frac{1}{1920 \pi^2}(N_s+6N_f+12 N_v),\\
  \lambda_2=\frac{1}{5760 \pi^2}(N_s+11N_f+62 N_v).
\end{align}
The first and the second part of eq.\eqref{cfano} are often called type B and type A anomaly in conformal field theory.

\section{The modified black hole geometry and thermodynamics}
Starting from the effective action \eqref{EFTaction}, the gravitational field equations can be derived by the its variation with respect to the metric, which can be put into the form
\begin{align}
R^{\mu \nu}-\frac{1}{2} R g^{\mu \nu}=8 \pi T^{\mu \nu}, \label{eins}
\end{align}
where $T^{\mu \nu}$ represents the quantum corrections to the classical vacuum, as stated in last section. In the following, our main aim is to solve the corrections to the original Schwarzschild geometry from the field equations \eqref{eins}, and analyze the modified black hole thermodynamics.

\subsection{Quntum corrections to the Schwarzschild metric}
The difficulty is that the effective action is complicated and the non-local operator $\ln \Box$ is intricate. We split $ T^{\mu \nu}\equiv \frac{2}{\sqrt{g}} \frac{\delta}{\delta g_{\mu\nu}} I_q$ into two portions: $T^{\mu \nu}=H^{\mu \nu}+ K^{\mu \nu}$. Here $H^{\mu \nu}$ collects all the terms related to the variation of $ \sqrt{-g}$ and $R_{\mu\nu\alpha\beta}$ in the quantum part of the Lagrangian $\sqrt{g}\mathcal{L}_q$, which is written explicitly as
 \begin{align}
 H^{\mu \nu}\equiv\frac{2}{\sqrt{g}} \left( \frac{\delta \sqrt{g}}{\delta g_{\mu \nu}}\mathcal{L}_q +  \cdots  -  \sqrt{g}  2 \gamma \frac{ \delta{R_{\alpha\beta \rho \sigma}}  }{\delta{g_{\mu \nu}}}  \ln( \frac{\Box}{\mu^2} ) R^{\alpha\beta \rho \sigma}   \right),
 \end{align}
 wherein the ``$\cdots$" represent the contributions from other curvature terms in the Lagrangian, and $K^{\mu \nu}$ collects only the terms coming from $\delta \ln \Box$, that is,
 \begin{align}
K^{\mu \nu} \equiv -2 \left(\alpha R \frac{\delta{\ln \Box}}{\delta g_{\mu \nu}} R  + \beta R_{\mu\nu} \frac{\delta{\ln \Box}}{\delta g_{\mu \nu}} R^{\mu\nu}  + \gamma   R_{\alpha \beta\rho\sigma} \frac{\delta{\ln \Box}}{\delta g_{\mu \nu}} R^{\alpha \beta\rho\sigma} \right). \label{kuv}
\end{align}

  The portion $H^{\mu \nu}$ can be derived straightforwardly, though its form is rather lengthy \cite{Calmet:2019eof}. We are only interested in the corrections around the Schwarzschild black hole metric at the first order in the Wilson coefficients $\alpha$, $\beta$, $\gamma$, so a lot of terms in $H_{\mu \nu}$ vanish and it reduces to
\begin{align}  H^{\mu \nu}  =4  \gamma   ( \nabla_\alpha \nabla_\beta+ \nabla_\beta \nabla_\alpha  ) \ln(\frac{\Box}{\mu^2}) R^{\mu \alpha \nu \beta}. \label{huv}
\end{align}
A subtlety is that there has already been a coefficient $\gamma$ in eq.\eqref{huv}, so we only need to use the original Schwarzschild metric to evaluate it. Clearly $H^{\mu \nu}$ is traceless for all the Ricci-flat geometries
\begin{align}
    H^\mu_{\ \mu} \equiv g_{\mu \nu} H^{\mu \nu}= \gamma ( \nabla_\alpha \nabla_\beta+ \nabla_\beta \nabla_\alpha  ) \ln(\frac{\Box}{\mu^2}) R^{ \alpha\beta}=0.
\end{align}
Concretely speaking, the metric $g_{\mu \nu}$ can commute with the covariant derivative operators $\nabla_\mu$ and $\Box$ (notice $\ln \Box$ can be formally represented as the form $\ln \frac{\Box}{\mu^2} =\int_0^{\infty}ds(\frac{1}{\mu^2+s}-\frac{1}{\Box+s})$) and contract with  $R^{\mu \alpha \nu \beta}$, so the trace vanishes due to Ricci-flat condition $R^{ \alpha\beta}=0$.

At present, there is no available technique to deal with $\ln (\frac{\Box}{\mu^2}) R_{\mu \alpha \nu \beta}$ in a curved spacetime in evaluating eq.\eqref{huv}, but $-\ln(r^2 \mu^2)R_{\mu \alpha \nu \beta}$ would be the only conceivable result by dimensional argument, which is also supported by the calculation in flat spacetime \footnote{Using a formula in the appendix of \cite{Calmet:2019eof}, there is $\ln \frac{ \Box^{flat}}{\mu ^2} \frac{M}{r^3}=-( \ln (r^2 \mu^2) +2 \gamma_E-2 )\frac{M}{r^3}$ up to a divergent part which may be cured by introducing appropriate boundary conditions or some regularization scheme. Thus we naturally expect $\ln (\frac{\Box}{\mu^2}) R_{\mu \alpha \nu \beta}=-\ln(r^2 \mu^2)R_{\mu \alpha \nu \beta}$, after absorbing the constant coefficients $2 \gamma_E-2$ (usually related to the loop integration) into $c_3({\mu})$ for simplicity. Whatever, the main strategy is to do the calculation and check the self-consistencies of the results.}. Thus we obtain
\begin{align}
\begin{split}
    H^{\mu}_{\  \nu} &=\gamma\{(-\frac{32 M (5M-2r)}{r^6},0,0,0),(0,-\frac{32 M (3M-r)}{r^6},0,0),\\&(0,0,\frac{16 M (8M-3r)}{r^6},0),(0,0,0,\frac{16 M (8M-3r)}{r^6})\}.
    \end{split}
 \label{huv1}
\end{align}
One can easily check it is traceless as promised, by adding all the diagonal elements together.

The variation of $\ln \Box$ in eq.\eqref{kuv} is even more difficult to evaluate and no one knows how to calculate it directly and explicitly. However, it doesn't mean we know nothing about $K^{\mu \nu}$. For example, its trace should be the anomalous term as analyzed in Sec.\ref{sec2}, that is,
\begin{align}
     K^\mu_{\ \mu}= 2 (\alpha R^2 + \beta R_{\mu\nu} R^{\mu\nu} + \gamma R_{\mu\nu\alpha\beta}  R^{\mu\nu\alpha\beta})=\gamma \frac{96  M^2}{r^6}, \label{cond1}
\end{align}
 where in the last step we provided the value for the Schwarzschild case. On the other hand, since both $H^{\mu \nu}$ and $K^{\mu \nu}$ are generated from the variation of the same Lagrangian, they have to be correlated with each other. Especially, according to Bianchi Identity, there should be \begin{align}
  \nabla_\mu T^{\mu \nu}= \nabla_\mu ( H^{\mu \nu}+ K^{\mu \nu})=0,
 \end{align}
 which leads to
 \begin{align}
 \nabla_\mu  K^{\mu \nu}=-\nabla_\mu  H^{\mu \nu}=\{0,\frac{96   M^2 (2 M-r)\gamma}{r^8},0,0\}. \label{cond2}
 \end{align}
From the two conditions \eqref{cond1} and \eqref{cond2}, we get the form of $K^{\mu \nu}$ as
\begin{align}
      K^{\mu}_{\ \nu}=\{(0,0,0,0),(0,0,0,0),
    (0,0,\frac{48  M^2\gamma}{r^6},0),(0,0,0,\frac{48  M^2\gamma}{r^6})\}.
 \label{kuv3}
\end{align}
Actually, the above two conditions are not enough to determine the form of $K^{\mu \nu}$ completely, and one can still add a symmetric, divergence free and trace free tensor to it.
An explanation of what happens in adding such tensors and why we adhere to the choice of the form \eqref{kuv3} will be provided at the end of Sec.\ref{secmodther}, to avoid diverging too much from the main thread.

Then we can solve the gravitational field equations around the Schwarzschild metric under the ansatz
 \begin{align}
    ds^2=-f(r)dt^2 +\frac{1}{g(r)} dr^2 +r^2 d \Omega,\label{ansatz}
\end{align}
where $f(r)=1-\frac{2M}{r}+\gamma a(r)$, $g(r)=1-\frac{2M}{r}+ \gamma b(r)$. Substituting eqs.\eqref{huv1}, \eqref{kuv3} and \eqref{ansatz} into \eqref{eins} and only keeping the terms linear in $\gamma$, we get the $tt$ and $rr$ components of the gravitational field equations as
\begin{small}
\begin{align}
    & \frac{\left(b(r)+r b'(r)\right)\gamma}{r^2} =\frac{256 \pi M(2r-5M) \gamma}{r^6},\\
  & \frac{\left(-2M a(r) \! + r (r-2M) a'(r) \!+ r b(r)\right)\gamma}{r^2 (r-2M)}\!=\!\frac{256 \pi M(r-3M) \gamma}{r^6},
\end{align}
\end{small}
from which we find the solution
\begin{align}
\begin{split}
f(r)=1-\frac{2 M}{r}+ \gamma \hbar G (-\frac{512 \pi   M}{3r^3} + \frac{256 \pi   M^2}{r^4}),\\
 g(r)=1-\frac{2 M}{r}+\gamma \hbar G ( -\frac{256 \pi   M}{r^3} +  \frac{1280 \pi   M^2}{3 r^4}),
 \end{split}\label{2ndsolution}
\end{align}
 where have temporarily restored $\hbar G$ to highlight this is a quantum-gravitational correction to the metric. The other components of the field equations \eqref{eins} are satisfied automatically.

 By hindsight, actually we don't need to worry too much about the annoying $K^{\mu \nu}$ in the above procedure. An easier way is to solve the field equations directly from the $tt$ and $rr$ components of $H^{\mu \nu}$ (two independent components of the field equations are enough to solve the two unknown functions $a(r)$ and $b(r)$ in the modified metric), then with the derived metric, to check the trace of the Einstein tensor is the same as that required by the trace anomaly. Whatever, we have to stress that it is just a technical trick using solely $H^{\mu \nu}$ to solve the field equations. In principle, a consistent calculation requires to use the complete form of $T^{\mu \nu}=H^{\mu \nu}+K^{\mu \nu}$ to solve the field equations. The reason why such a trick works is because the $tt$ and $rr$ components of $K^{\mu \nu}$ vanish in this case (see eq.\eqref{kuv3}), then the two components of $H^{\mu \nu}$ equal to those of $T^{\mu \nu}$. Needless to say, unless adding $K^{\mu \nu}$ to it, the angular components of $H^{\mu \nu}$ will cause trouble in solving the field equations. In a word, we propose such a trick with the purpose to simplify the complicated calculation. In general cases, when a metric has been successfully solved, one must check the trace anomaly pops out correctly (by calculating the trace of Einstein tensor, and subtracting other possible sources of the nonzero trace). If not, a detailed analysis of $K^{\mu \nu}$ is unavoidable. We have further considered the R-N and AdS black holes (not Ricci-flat) and carried out higher-order calculations. It is an astonishing (unexpected) fact that the trick works even in these cases. This may imply some nice property of the operators $\ln \Box$ and $\delta \ln \Box$ underlying the intricate problems.

  By the way, there were some controversies around the quantum corrections to the black hole metrics from various motivations in the literature \cite{BjerrumBohr:2002ks,Kirilin:2006en,Calmet:2017qqa}. In particular, it was argued in a recent paper \cite{Calmet:2021lny} that $T_{\mu \nu}$ generated from the action \eqref{EFTaction} is only at $\mathcal{O}(R^3)$ or higher order in curvature, hence it is dropped off at second order. Accordingly, the gravitational field equations and the Schwarzschild metric receive no corrections there. The main distinction between the present work and \cite{Calmet:2021lny} is that $T_{\mu \nu}$ doesn't vanish from our own analysis. The differences may be fixed in future with more powerful tools being developed in dealing with the non-local operator $\ln \Box$ especially in curved space-times. At present, we emphasize that the feature and advantage of our result \eqref{2ndsolution} is that both the modified thermodynamics \eqref{cther} and the trace (conformal) anomaly \eqref{anom} can be realized simultaneously.

 \subsection{The modified black hole thermodynamics}\label{secmodther}

With the quantum corrections to the black hole metric \eqref{2ndsolution}, we can further analyze the modified thermodynamics. Requiring $f(r_h)=g(r_h)=0$, we find the radius of the horizon
 \begin{align}
r_h=2M+\gamma \frac{32 \pi}{3M},\label{rhh}
\end{align}
so we still have a well-defined horizon of radius $r_h$ on which the thermodynamics can be constructed. By analyzing the conical singularity of the metric, the black hole temperature is
\begin{align}
 T=\frac{\sqrt{ f'(r_h)g'(r_h)}}{4\pi} =\frac{1}{8\pi M}-\gamma \frac{2}{M^3}.
 \label{tem} \end{align}
We also verified it by calculating the surface gravity at the horizon. The black hole entropy can be derived from the Wald formula, which reads \cite{Wald:1993nt}
\begin{align}
S_{W} =  -2\pi \oint
	 (\frac{\partial \mathcal{L}}{\partial R_{\mu\nu\rho\sigma}})^{(0)} \epsilon_{\mu\nu} \epsilon_{\rho\sigma} d\Sigma,\label{wald}
\end{align}
where $d\Sigma=r^2 \sin \theta d\theta d\phi$, $\mathcal{L}$ is the Lagrangian of the theory, and $\epsilon_{\mu\nu}$ should be normalized with $\epsilon_{\mu\nu}\epsilon^{\mu\nu}=-2$ which means $\epsilon_{tr}=\sqrt{f(r)/g(r)} $ etc. Using the effective action \eqref{EFTaction} and the metric \eqref{2ndsolution}, we derive the entropy
\begin{align}
\begin{split}
S_W  &= \pi r_h^2 + 64 \pi^2 c_3(\mu) +64 \pi^2 \gamma \ln  ( r_h^2 \mu^2)\\
& = 4 \pi M^2  + 64 \pi^2 \gamma  \ln  ( 4 M^2 \mu^2 ) + 64 \pi^2 c_3(\mu) + \gamma\frac{128 \pi^2}{3}.
\end{split}\label{entropy01}
\end{align}
 The three terms in the first line respectively comes from the Einstein-Hilbert part, the other local part, and the non-local part of the EFT action. Note the entropy \eqref{entropy01} has already been renormalization group invariant, i.e., $\mu$-independent, and $c_3(\mu)+\gamma \ln(\mu^2)$ should correspond to some characteristic scale $l_s$ of the complete quantum gravity \cite{El-Menoufi:2015cqw}. If one believes no new scales other than the Planck scale $l_p$, eq.\eqref{entropy01} would be written as a neat form $S_{W}  =   \frac{\mathcal{A}}{4 l_p^2} + 64 \pi^2 \gamma \ln  ( \frac{\mathcal{A}}{l_p^2} )$.

 Now we have found the quantum corrections to the original Schwarzshild metric, black hole temperature and entropy at the first order in the Wilson coefficients. Next we illustrate that the modified black hole geometry and thermodynamics are self-consistent. First, we can check easily  from eqs.\eqref{tem} and \eqref{entropy01} that the thermodynamic equation $TdS=dM$ holds. Second, we check the Euler characteristic is still $\chi=2+\mathcal{O}(\gamma^2)$, that is,
 \begin{equation}
	\chi
	= \frac{1}{32\pi^2} \int_{0}^{1/T} dt_E \int_{r_H}^{\infty} dr d\theta d\phi
	\sqrt{-g} \left(R^2 - 4 R_{\mu\nu} R^{\mu\nu} + R_{\mu\nu\rho\sigma} R^{\mu\nu\rho\sigma} \right)=2. \label{eulerxi}
	\end{equation}
In the above expression, we used all the corrected values of $\sqrt{-g}$, curvature tensors, the horizon radius and the temperature. It is amazing to see all these complicated numbers interacting with each other to produce $0$ at first order in $\gamma$. This is a rather stringent examination, since that if we got any of these quantities incorrect, $\chi=2$ couldn't come out. Third, we redone the calculation \cite{El-Menoufi:2015cqw,El-Menoufi:2017kew} using the Euclidean action approach. Now the Euclidean action is evaluated and explained as the thermodynamic partition function
\begin{align}
I_{EFT} = -\frac{\beta ^2}{16 \pi } +64 \pi ^2 \gamma  \ln \left(\frac{\beta ^2 \mu ^2}{16 \pi ^2}\right)+\frac{128 \pi ^2 \gamma }{3},
 \end{align}
 where $\beta\equiv 1/T$. Then the black hole mass can be calculated by $M(\beta)=-\frac{\partial}{\partial \beta} I_{EFT}$, which conversely leads to the relation between the black hole temperature and mass
 \begin{align}
   T=   \frac{1}{8\pi M}-\gamma \frac{2}{M^3}. \label{temeu}
 \end{align}
Then the Euclidean entropy is
\begin{align}
S_E = \beta M + I_{EFT} = 4 \pi M^2  + 64 \pi^2 \gamma  \ln  ( 4 M^2 \mu^2 ) + 64 \pi^2 c_3(\mu) + \gamma\frac{128 \pi^2}{3}. \label{entreu}
\end{align}
The Euclidean entropy exactly matches with the Wald entropy \eqref{entropy01} including the constant term $\gamma\frac{128 \pi^2}{3}$. The temperature \eqref{temeu} is also the same as eq.\eqref{tem}, and implies a modified geometry from the original Schwarzschild metric.

The thermodynamic behaviors \eqref{tem} and \eqref{entropy01} and their variants have been given in the earlier literature \cite{Solodukhin:2011gn,Solodukhin:1994yz,Fursaev:1994te,Sen:2012dw,El-Menoufi:2015cqw,El-Menoufi:2017kew}. However, using the EFT formalism, we have also provided the quantum corrections to the black hole geometry, accompanied by a fully consistent analysis. The higher order corrections to the black hole thermodynamics caused by the effective action can be computed as well \footnote{As an example, at 2nd order in the Wilson coefficients, we derive
$T =\frac{1}{8\pi M}-\gamma \frac{2}{M^3} + \gamma^2 \frac{736 \pi}{15 M^5}$ and
   $S = 4 \pi M^2 + 64 \pi^2 c_3(\mu) +64 \pi^2 \gamma  \ln  ( 4 M^2 \mu^2 )  + \gamma\frac{128 \pi^2}{3}  +  \gamma^2 \frac{8192 \pi^3}{15 M^2}$.
In fact, the metric has not been Ricci-flat at the 1st order, so the parameters $\alpha$, $\beta$ must be taken into consideration in computing the higher order results. They are not present explicitly in the above expressions only because they cancelled out using $M$ as the variable.}.

Finally, we come to discuss the freedom of choosing $K^{\mu}_{\ \nu}$ as mentioned. The form of $K^{\mu}_{\ \nu}$ in eq.\eqref{kuv3} is determined up to adding a symmetric, divergence free and trace free tensor to it. Below we make two comments on this issue. First, adding such tensors will not change the main thermodynamic behaviors \eqref{cther} that we intend to reproduce. For instance, we add to eq.\eqref{kuv3} an extra divergence free and trace free tensor
\begin{align}
\begin{split}
c_1  \{ ( \frac{M (7 M - 4 r) \gamma}{r^6}, 0, 0, 0 ), ( 0, \frac{M (3 M - 2 r)\gamma }{r^6}, 0, 0 ),
(0, 0, \frac{M (-5 M + 3 r) \gamma}{r^6}, 0 ), (0, 0, 0, \frac{M (-5 M + 3 r) \gamma}{r^6} \},
   \end{split}
   \end{align}
where $c_1$ is a free parameter. Applying the alternative choice of $K^{\mu}_{\ \nu}$, the solved metric has to be modified accordingly, and the horizon radius becomes $r_h=2M+\gamma \frac{32 \pi}{3M}-\gamma \frac{5 \pi  }{3M}c_1$ compared to eq.\eqref{rhh}. Substituting it into the first line of the Wald entropy \eqref{entropy01}, the first term gives $\pi r_h^2 =4 \pi M^2 +  \gamma \frac{128 \pi^2}{3} -  \gamma \frac{20 \pi^2}{3} c_1 +\mathcal{O}(\gamma^2) $, and the last two terms do not change at this order. So it only brings about an extra constant term $-  \gamma \frac{20 \pi^2}{3} c_1$ to the entropy. The temperature can be calculated straightforwardly from the corresponding metric, which is still the form \eqref{tem}. This is understandable since $TdS=dM$ must hold if the calculation is executed correctly. Though the above analysis is for a concrete example, the conclusion doesn't change by adding more complicated divergence free and trace free tensors. The key point is that $\gamma$ is dimensionless, so the extra shift to the horizon radius is destined to be proportional to $\gamma \hbar G /M$, which only causes an extra constant term to the entropy. Obviously, this is a specialty of the present theory at hand. For cubic gravity or higher order gravity where the coupling coefficients are dimensional, adding such tensors to the original energy-momentum tensor should modify the black hole thermodynamics more obviously. Second, we explain why we adhere to the choice of eq.\eqref{kuv3}. The Euclidean integral approach is an independent way to deal with black hole thermodynamics, and generally one expects an agreement between the Wald entropy and the Euclidean entropy  \cite{Dutta:2006vs}. Below eq.\eqref{entreu}, we have emphasized such a nice agreement between eq.\eqref{entropy01} and eq.\eqref{entreu}. In a word, we don't have a strict proof of choosing eq.\eqref{kuv3} as $K^{\mu}_{\ \nu}$, but fixed it by the requirement of consistency. It turns to be that the simplest choice is the most consistent one.

\section{Generalizations to other types of black hole}

Our formalism can be straightforwardly generalized to study the quantum gravitational corrections to the geometries and thermodynamics of other types of black hole. In this section, we will consider the R-N black hole and AdS-Schwarzschild black hole as the examples. The calculations are more complicated and involved, but the procedures are much alike, so we only list the results as below. Notice that we will omit the local Wilson coefficients $c_1(\mu)$, $c_2(\mu)$ and $c_3(\mu)$ to simplify the expressions. One can easily retrieve them because they always come together with $\alpha \ln (\mu^2)$, $\beta \ln (\mu^2)$ and $\gamma \ln (\mu^2)$; and upon doing this, all the results are renormalization group invariant, i.e., $\mu$-independent, as explained earlier.

\subsection{Quantum corrections to the R-N black hole} \label{rnsec}
Adding an electromagnetic field part to the effective action \eqref{EFTaction}, around a R-N black hole geometry, the metric with quantum corrections can be solved as
\begin{equation}
\begin{split}
f(r)=& 1-\frac{2M}{r}+\frac{q^2}{r^2} - \frac{32 \pi}{75 r^6}\left[ 200 M r^2 (-3M+2r) \gamma + 25 q^2 (2Mr-9r^2) \gamma + 3q^4 (2\beta-47 \gamma) \right. \\  & \left.+15q^2 (q^2-5Mr+5 r^2)(\beta+4\gamma) \ln (r^2 \mu^2) \right],\\
g(r)=& 1-\frac{2M}{r}+\frac{q^2}{r^2}+\frac{32 \pi}{75 r^6}\left[ 200Mr^2(5M-3r) \gamma +75q^2 r (-4M \beta+2r \beta-29M \gamma+12 r\gamma)\right. \\ & \left.
+48q^4 (3\beta+17\gamma)-15 q^2 (6q^2 -15Mr+10r^2)(\beta+4\gamma) \ln (r^2 \mu^2) \right].
\end{split}
\end{equation}
The relation between the mass $M$ and the outer horizon radius $r_+$ of the R-N black hole can be obtained by requiring $f(r_+)=g(r_+)=0$, which gives
\begin{align}
\begin{split}
M=&\frac{r_+^2+q^2}{2r_+} + \frac{8\pi}{75 r_+^5} \left[ -100 r_+^4 \gamma + 25 q^2 r_+^2 \gamma -q^4 (12\beta+43 \gamma)\right. \\  & \left.
+15 q^2 (3q^2-5 r_+^2) (\beta+4\gamma)\ln (r_+^2 \mu^2) \right].
\end{split}
\end{align}
Obviously, it will be more convenient to use $r_+$ instead of $M$ as the variable to do the analysis. The electrostatic potential at the horizon is
\begin{align}
   \Phi = \frac{q}{r_+} +\frac{16\pi q}{75 r_+^5 } \left[25 r_+^2 \gamma-2q^2 (12\beta+43\gamma)+ 15 q^2 (\beta+4\gamma) \ln(r_+^2\mu^2) \right].
\end{align}
The modified temperature is calculated by analyzing the conical singularity of the metric
\begin{align}
    T=\frac{1}{4\pi r_+} -\frac{q^2}{4\pi r_+^3}-\frac{4 (r_+^2-q^2)}{3 r_+^7} \left[ (8r_+^2-q^2)\gamma -3q^2 (\beta+4\gamma)  \ln(r_+^2\mu^2)\right] .
\end{align}
The modified entropy is calculated by the Wald entropy formula
\begin{align}
    S_W=\pi r_+^2 + \frac{32\pi^2 }{r_+^2 } \left[ 2r_+^2 \gamma-q^2 (\beta+4\gamma)\right] \ln(r_+^2\mu^2) .\label{rnen}
\end{align}
Though there are always the weird $\ln r_+$ terms rambling around in the above expressions, the analysis of the thermodynamics goes smoothly. The complete calculation is tedious, but the reader can easily check the thermodynamic law $Td S+ \Phi dq=dM$ holds using $r_+$ and $q$ as the independent variables.

A known result for the R-N black hole coupled with massless scalar field is $S_{RN}=\frac{\mathcal{A}}{48 \pi \epsilon^2}+\frac{2r_{+}-3r_{-}}{90 r_+}\ln \frac{r_+}{\epsilon}$ where $\epsilon$ is an ultraviolet cutoff, in the language of entanglement entropy. And in the limit $r_-\rightarrow r_+$, it gives the entropy for the extremal case \cite{Solodukhin:2011gn}. When dealing with the Schwarzschild black hole, one can derive the expression of temperature $T(M)$ from that of entropy or vice versa, because of $TdS=dM$ \cite{Fursaev:1994te,Solodukhin:2019xwx}. This doesn't work for the R-N black hole due to the presence of the extra term $\Phi dq$. So the expressions for the thermodynamic quantities other than the entropy $S_{RN}$ were simply absent in the literature.  For comparison, setting $N_s=1$, $N_f=N_v=N_g=0$ (solely a scalar field) in eqs.\eqref{a1}--\eqref{gamma}, our result \eqref{rnen} soon reduces to a rather similar form $S_{RN}=\frac{\mathcal{A}}{4 \l_p^2}+\frac{2r_{+}-3r_{-}}{90 r_+}\ln (r_+ \mu)$. In addition, all the thermodynamic quantities have been obtained independently and the thermodynamic law $Td S+ \Phi dq=dM$ is only used for checking the consistency.

\subsection{Quantum corrections to the AdS-Schwarzschild black hole}
 In this case, the classical part of the effective action \eqref{EFTaction} becomes $S= \int_{\mathcal{M}} \frac{1}{16 \pi} (R-2\Lambda)$. The methods of analyzing the thermodynamics of an AdS-Schwarzschild black hole often require a procedure of subtracting a pure AdS geometry as the background. Therefore, in order to avoid unnecessary subtleties, we keep the pure AdS geometry still being a saddle point (solution) of the total effective action, and this can be fulfilled under the constraint
\begin{align}
    12\alpha+3 \beta+2\gamma=0. \label{constaint}
\end{align}
Notice it includes the case of type B anomaly where $\alpha=\frac{1}{3}$, $\beta=-2$, $\gamma=1$, see eq.\eqref{cfano}. Applying the constraint \eqref{constaint} to the effective action and solving the corresponding gravitational field equations, direct calculation gives the metric
\begin{align}
\begin{split}
& f(r)=1-\frac{2M}{r} -\frac{\Lambda r^2}{3} -\gamma \frac{c_0 M }{r}-\gamma  \frac{256 \pi M}{9 r^4}(-9M+6r+r^3 \Lambda +3  \Lambda r^3 \ln r),\\
&g(r)=1-\frac{2M}{r} -\frac{\Lambda r^2}{3} -\gamma \frac{c_0 M }{r} -\gamma  \frac{256 \pi M}{3 r^4}(-5M+3r+  \Lambda r^3\ln r),
\end{split} \label{adsgeo}
\end{align}
where we take $\Lambda<0$ and $c_0$ is a constant of integration. The relation between the mass $M$ and the outer horizon radius $r_+$ can be obtained by requiring $f(r_+)=g(r_+)=0$, which gives
\begin{align}
     M=\frac{(3-\Lambda r_+^2)r_+}{6}  - \gamma \frac{1}{108 r_+}(3- \Lambda r_+^2)\left( 9c_0 r_+^2 +128 \pi (3+5 \Lambda r_+^2)+ 768 \pi r_+^2 \Lambda \ln r_+ \right).
\end{align}
The modified temperature is calculated by analyzing the conical singularity of the metric
\begin{align}
    T=\frac{1}{4\pi r_+}-\frac{\Lambda r_+}{4 \pi} - \gamma \frac{32}{9 r_+^3} (3- 4 \Lambda r_+^2 + \Lambda^2 r_+^4 ).
\end{align}
The modified entropy is calculated by the Wald entropy formula
\begin{align}
    S_W=\pi r_+^2 + \gamma \frac{64 \pi^2}{3} (3- \Lambda r_+^2) \ln (r_+^2 \mu^2)  .
\end{align}
The thermodynamic law $TdS=dM$ fixes the constant $c_0=\frac{128 \pi \Lambda}{3}(-2+\ln (\mu^2))$. Substituting it into eq.\eqref{adsgeo}, the form of the metric can be improved as
\begin{align}
   \begin{split}
& f(r)=1-\frac{2M}{r} -\frac{\Lambda r^2}{3} + \gamma  \frac{128 \pi M}{9 r^4}(18M-12 r + 4 r^3 \Lambda - 3   \Lambda r^3 \ln (r^2 \mu^2)),\\
&g(r)=1-\frac{2M}{r} -\frac{\Lambda r^2}{3} +\gamma  \frac{128 \pi M}{3 r^4}(10M-6r+2 \Lambda r^3-  \Lambda r^3\ln (r^2 \mu^2)).
\end{split} \label{adsgeo2}
\end{align}

To verify the consistency, we further analyze the thermodynamics of the modified AdS-Schwarzschild black hole using the standard Euclidean action approach \cite{hawkingpage,Dutta:2006vs}. Evaluating the effective action at the saddle point \eqref{adsgeo2}, and cancelling its divergences by subtracting the contribution from the pure AdS geometry, the Euclidean action can be obtained and explained as the partition function. We find the thermodynamic quantities from the Euclidean approach exactly matches with the above formulas.

\section{Conclusions}

In this paper, we have solved the quantum corrections to the black hole geometries and thermodynamics from the one-loop effective action of quantum gravity, with the Schwarzschild, R-N and AdS-Schwarzschild black holes as the examples. In particular, the logarithmic corrections to the black hole entropy have been obtained and the consistencies of the thermodynamics have been examined. Our work suggests that the EFT approach provides a powerful and self-consistent tool for studying the quantum corrections even for more complicated types of black holes and to higher orders in the perturbation theory of quantum gravity. The three main progresses of the paper are summarized as below.

First, we have identified the \emph{non-local} effective action that is related to the logarithmic correction to the black hole entropy, which was not clear in most previous literature. Interestingly, the nonlocal term in \eqref{EFTaction} are not added by hand, they naturally emerge from integrating out the massless sector of the particle spectrum. In fact, the nature has provided another clues for such a nonlocality. As mentioned in the Introduction, this logarithmic correction can not be derived from \emph{local} curvature terms added to the effective action. Schematically, if one starts from an action of the form $I=\int_\mathcal{M} R+ R_{\mu\nu\rho\sigma} ^2 + R_{\mu\nu\rho\sigma}^3 + \cdots$, the black hole temperature will be of the form $T=\frac{1}{8\pi M} + \frac{1}{M^5}+\frac{1}{M^7}+\cdots$. It seems that there is a ``mysterious'' missing piece proportional to $\frac{1}{M^3}$ in this expression \cite{Xiao:2021ewv}. Disturbed for a long time by the faith of the perfection of the mathematical form, one may choose to argue for the existence of the $\frac{1}{M^3}$ term in the temperature, which should be attached to the $R_{\mu\nu\rho\sigma} (\ln \Box) R^{\mu\nu\rho\sigma}$ correction to the Einstein-Hilbert action according to our analysis. So this may serve as an alternative way to discover the non-local operator $\ln \Box$.

Second, we have solved the quantum correction to the geometry of the Schwarzschild black hole, which naturally leads to the modified thermodynamics \eqref{cther} and the anomalous trace \eqref{anom}. This could be a good complement to the previous scaling analysis by Fursaev in \cite{Fursaev:1994te}.

Third, we have further obtained the results for R-N black hole and AdS-Schwarzschild black hole. Because the modified geometries were unknown before, there were only limited results on the corrections to the entropy (see the last paragraph of Sec.\ref{rnsec}). In contrast, all thermodynamic quantities can be calculated consistently from our approach, which accommodates previously known results as specific examples.

At this stage, we only considered the quantum corrections to the static spherically symmetric black holes, so the rotating or other types of black hole can be analyzed in future. It was suggested that the quantum aspects of gravity could have observational results, for example, in the gravitational wave signals \cite{Laghi:2020rgl} and the black hole shadows \cite{Peng:2020wun}. In addition, our formalism may be also useful in studying the conformal anomaly for black holes in even dimensions other than $4$. And from the spirit of AdS/CFT, it is worthy to explore the meaning of the modified black hole geometry at the CFT side \cite{Azeyanagi:2009wf}. In cosmology, conformal anomaly could provide an explanation for the cosmological constant problem and has experimental predictions \cite{Thomas:2009uh,Antoniadis:2006wq}.

\section*{Acknowledgments}
YX would like to thank X. Calmet and F. Kuipers for extensive discussions, and is grateful to MPS School of the University of Sussex for the research facilities and the hospitality during the one-year visit, where this research was initiated and developed greatly. This work was supported in part by China Scholarship Council with Grant No. 201908130079 and NSFC with Grant No. 11975235 and 12035016, and NSF of Hebei province with Grant No. A2021201022.

%%%%%%%%%%%%%%%%%%%%%%%%%%%%%%%%%%%%%%%%%%%%%%%%%%%%%%%%%%%%%%%%%
%%%
%%%                     BIBLIOGRAPHY
%%%
%%%%%%%%%%%%%%%%%%%%%%%%%%%%%%%%%%%%%%%%%%%%%%%%%%%%%%%%%%%%%%%%%

\bigskip{}


\baselineskip=1.6pt

\begin{thebibliography}{10}

\bibitem{Maldacena:1997de}
J.~M.~Maldacena, A.~Strominger and E.~Witten,
\emph{Black hole entropy in M theory,}
JHEP \textbf{12}, 002 (1997)
[arXiv:hep-th/9711053 [hep-th]].

%\cite{Solodukhin:2011gn}
\bibitem{Solodukhin:2011gn}
S.~N.~Solodukhin,
\emph{Entanglement entropy of black holes,}
Living Rev. Rel. \textbf{14} (2011), 8
[arXiv:1104.3712 [hep-th]].
%297 citations counted in INSPIRE as of 15 Jul 2020

%\cite{Solodukhin:1994yz}
\bibitem{Solodukhin:1994yz}
S.~N.~Solodukhin,
\emph{The Conical singularity and quantum corrections to entropy of black hole,}
Phys. Rev. D \textbf{51}, 609-617 (1995)
[arXiv:hep-th/9407001 [hep-th]].
%228 citations counted in INSPIRE as of 20 Nov 2020

%\cite{Fursaev:1994te}
\bibitem{Fursaev:1994te}
D.~V.~Fursaev,
\emph{Temperature and entropy of a quantum black hole and conformal anomaly,}
Phys. Rev. D \textbf{51}, 5352-5355 (1995)
[arXiv:hep-th/9412161 [hep-th]].
%155 citations counted in INSPIRE as of 15 Nov 2020

%\cite{Sen:2012dw}
\bibitem{Sen:2012dw}
A.~Sen,
\emph{Logarithmic Corrections to Schwarzschild and Other Non-extremal Black Hole Entropy in Different Dimensions,}
JHEP \textbf{04}, 156 (2013)
[arXiv:1205.0971 [hep-th]].
%130 citations counted in INSPIRE as of 10 Aug 2020



%\cite{El-Menoufi:2015cqw}
\bibitem{El-Menoufi:2015cqw}
B.~K.~El-Menoufi,
\emph{Quantum gravity of Kerr-Schild spacetimes and the logarithmic correction to Schwarzschild black hole entropy,}
JHEP \textbf{05} (2016), 035
[arXiv:1511.08816 [hep-th]].
%16 citations counted in INSPIRE as of 05 Jul 2020


%\cite{El-Menoufi:2017kew}
\bibitem{El-Menoufi:2017kew}
B.~K.~El-Menoufi,
\emph{Quantum gravity effects on the thermodynamic stability of 4D Schwarzschild black hole,}
JHEP \textbf{08} (2017), 068
[arXiv:1703.10178 [gr-qc]].
%3 citations counted in INSPIRE as of 05 Jul 2020

\bibitem{Cai:2009ua}
R.~G.~Cai, L.~M.~Cao and N.~Ohta,
\emph{Black Holes in Gravity with Conformal Anomaly and Logarithmic Term in Black Hole Entropy,}
JHEP \textbf{04}, 082 (2010)
[arXiv:0911.4379 [hep-th]].

%\cite{Carlip:2000nv}
\bibitem{Carlip:2000nv}
S.~Carlip,
\emph{Logarithmic corrections to black hole entropy from the Cardy formula,}
Class. Quant. Grav. \textbf{17}, 4175-4186 (2000)
[arXiv:gr-qc/0005017 [gr-qc]].


%\cite{Banerjee:2008cf}
\bibitem{Banerjee:2008cf}
R.~Banerjee and B.~R.~Majhi,
\emph{Quantum Tunneling Beyond Semiclassical Approximation,}
JHEP \textbf{06}, 095 (2008)
[arXiv:0805.2220 [hep-th]].
%299 citations counted in INSPIRE as of 19 Nov 2020

%\cite{Banerjee:2008fz}
\bibitem{Banerjee:2008fz}
\emph{Quantum Tunneling, Trace Anomaly and Effective Metric,}
Phys. Lett. B \textbf{674}, 218-222 (2009)
[arXiv:0808.3688 [hep-th]].
%127 citations counted in INSPIRE as of 19 Nov 2020

\bibitem{kaul}
R.~K.~Kaul and P.~Majumdar,
\emph{Logarithmic correction to the Bekenstein-Hawking entropy,}
Phys. Rev. Lett. \textbf{84}, 5255-5257 (2000)
[arXiv:gr-qc/0002040 [gr-qc]].

%\cite{Solodukhin:2019xwx}
\bibitem{Solodukhin:2019xwx}
S.~N.~Solodukhin,
\emph{Logarithmic terms in entropy of Schwarzschild black holes in higher loops,}
Phys. Lett. B \textbf{802} (2020), 135235
[arXiv:1907.07916 [hep-th]].
%1 citations counted in INSPIRE as of 17 Jul 2020

%\cite{Solodukhin:1997yy}
\bibitem{Solodukhin:1997yy}
S.~N.~Solodukhin,
\emph{Entropy of Schwarzschild black hole and string - black hole correspondence,}
Phys. Rev. D \textbf{57}, 2410-2414 (1998)
[arXiv:hep-th/9701106 [hep-th]].
%105 citations counted in INSPIRE as of 19 Nov 2020

%\cite{Donoghue:1994dn}
\bibitem{Donoghue:1994dn}
J.~F.~Donoghue,
\emph{General relativity as an effective field theory: The leading quantum corrections,}
Phys. Rev. D \textbf{50}, 3874-3888 (1994)
[arXiv:gr-qc/9405057 [gr-qc]].
%851 citations counted in INSPIRE as of 19 Nov 2020


%\cite{Donoghue:2014yha}
\bibitem{Donoghue:2014yha}
J.~F.~Donoghue and B.~K.~El-Menoufi,
\emph{Nonlocal quantum effects in cosmology: Quantum memory, nonlocal FLRW equations, and singularity avoidance,}
Phys. Rev. D \textbf{89}, no.10, 104062 (2014)
[arXiv:1402.3252 [gr-qc]].
%67 citations counted in INSPIRE as of 19 Nov 2020

%\cite{Barvinsky:1985an}
\bibitem{Barvinsky:1985an}
A.~O.~Barvinsky and G.~A.~Vilkovisky,
\emph{The Generalized Schwinger-Dewitt Technique in Gauge Theories and Quantum Gravity,}
Phys. Rept. \textbf{119}, 1-74 (1985)
%564 citations counted in INSPIRE as of 08 Apr 2021

%\cite{Calmet:2019eof}
\bibitem{Calmet:2019eof}
X.~Calmet, R.~Casadio and F.~Kuipers,
\emph{Quantum Gravitational Corrections to a Star Metric and the Black Hole Limit,}
Phys. Rev. D \textbf{100} (2019) no.8, 086010
[arXiv:1909.13277 [hep-th]].
%4 citations counted in INSPIRE as of 05 Jul 2020

%\cite{Calmet:2021lny}
\bibitem{Calmet:2021lny}
X.~Calmet and F.~Kuipers,
\emph{Quantum gravitational corrections to the entropy of a Schwarzschild black hole,}
Phys. Rev. D \textbf{104}, no.6, 066012 (2021)
[arXiv:2108.06824 [hep-th]].
%2 citations counted in INSPIRE as of 22 Jan 2022


%\cite{Teixeira:2020kew}
\bibitem{Teixeira:2020kew}
P.~d.~Teixeira, I.~L.~Shapiro and T.~G.~Ribeiro,
\emph{One-loop effective action: nonlocal form factors and renormalization group,}
Grav. Cosmol. \textbf{26}, no.3, 185-199 (2020)
[arXiv:2003.04503 [hep-th]].
%5 citations counted in INSPIRE as of 11 Apr 2021

\bibitem{qfcs} N.~D.~Birrell and P.~C.~W.~Davies, \emph{Quantum Fields in Curved Space,} Cambridge University Press, Cambridge, 1982.

%\cite{Duff:2020dqb}
\bibitem{Duff:2020dqb}
M.~J.~Duff,
\emph{Weyl, Pontryagin, Euler, Eguchi and Freund,}
J. Phys. A \textbf{53}, no.30, 301001 (2020)
[arXiv:2003.02688 [hep-th]].
%2 citations counted in INSPIRE as of 09 Apr 2021



%\cite{BjerrumBohr:2002ks}
\bibitem{BjerrumBohr:2002ks}
N.~E.~J.~Bjerrum-Bohr, J.~F.~Donoghue and B.~R.~Holstein,
\emph{Quantum corrections to the Schwarzschild and Kerr metrics,}
Phys. Rev. D \textbf{68}, 084005 (2003)
[erratum: Phys. Rev. D \textbf{71}, 069904 (2005)]
[arXiv:hep-th/0211071 [hep-th]].

%\cite{Kirilin:2006en}
\bibitem{Kirilin:2006en}
G.~G.~Kirilin,
\emph{Quantum corrections to the Schwarzschild metric and reparametrization transformations,}
Phys. Rev. D \textbf{75}, 108501 (2007)
[arXiv:gr-qc/0601020 [gr-qc]].

%\cite{Calmet:2017qqa}
\bibitem{Calmet:2017qqa}
X.~Calmet and B.~K.~El-Menoufi,
\emph{Quantum Corrections to Schwarzschild Black Hole,}
Eur. Phys. J. C \textbf{77}, no.4, 243 (2017)
[arXiv:1704.00261 [hep-th]].






%\cite{Wald:1993nt}
\bibitem{Wald:1993nt}
R.~M.~Wald,
\emph{Black hole entropy is the Noether charge,}
Phys. Rev. D \textbf{48} (1993) no.8, 3427-3431
[arXiv:gr-qc/9307038 [gr-qc]].
%1624 citations counted in INSPIRE as of 15 Jul 2020

\bibitem{hawkingpage}
S.~W.~Hawking and D.~N.~Page, \emph{Thermodynamics Of Black Holes In Anti-De Sitter
Space,} Commun. Math. Phys. 87, 577 (1983).


%\cite{Dutta:2006vs}
\bibitem{Dutta:2006vs}
S.~Dutta and R.~Gopakumar,
\emph{On Euclidean and Noetherian entropies in AdS space,}
Phys. Rev. D \textbf{74}, 044007 (2006)
[arXiv:hep-th/0604070 [hep-th]].
%35 citations counted in INSPIRE as of 20 Apr 2021

\bibitem{Xiao:2021ewv}
Y.~Xiao, Y.~Chen, H.~Feng and C.~Zhu,
\emph{Black hole solutions and thermodynamics in the infinite derivative theory of gravity,}
Phys. Rev. D \textbf{103}, no.4, 044064 (2021)


\bibitem{Laghi:2020rgl}
D.~Laghi, G.~Carullo, J.~Veitch and W.~D.~Pozzo,
\emph{Quantum black hole spectroscopy: probing the quantum nature of the black hole area using LIGO-Virgo ringdown detections,}
Class. Quant. Grav. \textbf{38}, no.9, 095005 (2021)
[arXiv:2011.03816 [gr-qc]].

%\cite{Peng:2020wun}
\bibitem{Peng:2020wun}
J.~Peng, M.~Guo and X.~H.~Feng,
\emph{Influence of Quantum Correction on the Black Hole Shadows, Photon Rings and Lensing Rings,}
[arXiv:2008.00657 [gr-qc]].


%\cite{Azeyanagi:2009wf}
\bibitem{Azeyanagi:2009wf}
T.~Azeyanagi, G.~Compere, N.~Ogawa, Y.~Tachikawa and S.~Terashima,
\emph{Higher-Derivative Corrections to the Asymptotic Virasoro Symmetry of 4d Extremal Black Holes,}
Prog. Theor. Phys. \textbf{122}, 355-384 (2009)
[arXiv:0903.4176 [hep-th]].


%\cite{Thomas:2009uh}
\bibitem{Thomas:2009uh}
E.~C.~Thomas, F.~R.~Urban and A.~R.~Zhitnitsky,
\emph{The Cosmological constant as a manifestation of the conformal anomaly?}
JHEP \textbf{08}, 043 (2009)
[arXiv:0904.3779 [gr-qc]].

%\cite{Antoniadis:2006wq}
\bibitem{Antoniadis:2006wq}
I.~Antoniadis, P.~O.~Mazur and E.~Mottola,
\emph{Cosmological dark energy: Prospects for a dynamical theory,}
New J. Phys. \textbf{9}, 11 (2007)
[arXiv:gr-qc/0612068 [gr-qc]].

\end{thebibliography}
\end{document}